# A chaotic long-lived vortex at Venus' southern pole

I. Garate-Lopez* [1], R. Hueso[1,2], A. Sánchez-Lavega[1,2], J. Peralta[3], G. Piccioni[4] and P. Drossart[5]

Corresponding autor: itziar.garate@ehu.eus (I. Garate-Lopez)

[1] Dpto. Física Aplicada I, Escuela Técnica Superior de Ingeniería, Universidad del País Vasco (UPV/EHU), Alda. Urquijo s/n, 48013, BILBAO (Spain)

[2] Unidad Asociada Grupo Ciencias Planetarias UPV/EHU-IAA(CSIC).

[3] CAAUL / Observatório Astronómico de Lisboa, Tapada da Ajuda, 1349-018 Lisboa, Portugal

[4] IASF / INAF, 100 Via del Fosso del Cavaliere, Rome, Italy

[5] LESIA / Observatoire de Paris, CNRS UPMC, Univ. Paris-Diderot, 5, Place Jules Janssen, 92195 Meudon, France

**Polar vortices are common in the atmospheres of rapidly rotating planets [1-4]. On Earth and Mars they are tied to the surface and their existence follows the seasonal insolation cycle [1-3]. Venus is a slowly rotating planet but it is also known to have vortices at both poles at the edge of a superrotating atmosphere [5-8]. However, their nature and long-term properties have not been constrained so far impeding precise modeling. Here we report cloud motions at two altitude levels (about 42 km and 63 km above the surface) using infrared images from the VIRTIS instrument onboard Venus Express that show that the south polar vortex is a permanent but erratic and unpredictable feature. We find that the centers of rotation of the vortex at these levels rarely coincide and both wander erratically around the pole with speeds of up to 16 m s$^{-1}$. The cloud morphology and vorticity patches are uncorrelated and change continuously developing transient areas of small vertical motions. Venus' south polar vortex is a continuously evolving structure immersed in a baroclinic environment laying at altitude levels that have variable vertical and meridional wind shears, extending at least 20 km in height through a quasi-convective turbulent region.**

We have used high-resolution images in the infrared range (1 – 5 µm) taken by the VIRTIS-M instrument onboard Venus Express (VEX) [9] to examine Venus' South Polar Vortex [8]. At 1.74 µm the thermal radiation from the lower atmosphere is filtered by the deep clouds forming cloud opacity images (at ~ 42 km altitude in polar latitudes [10]). At 3.80 and 5.10 µm we observe thermal emission images from the upper cloud (at ~ 63 km altitude in polar latitudes [11, 12]) with different contrast and different capability to simultaneously show the polar day- and night-sides. The vortex is a permanent feature observed by VIRTIS-M infrared channel from the Venus Orbit Insertion (VOI) in April 2006 to October 2008, and beyond that period by the VIRTIS-M visible channel and the Venus Monitoring Camera, albeit with less contrast [13]. Images of the polar vortex at 3.80 or 5.10 µm over 169 orbits (1 orbit = 24 hr) show that in 35% of the orbits the vortex presents a dipolar shape (Fig. 1). In other cases it appears as an elongated oval (25%) or with a nearly circular structure (10%). Most of the other observations show the vortex as a transition feature between these configurations with an irregular structure in some cases (see supplementary information). The vortex morphology can be stable over tens of days (it was observed as a dipole from VOI to orbit 38) [8] or change on the time-scales of just 1 day [14]. In those cases where the vortex preserves an identifiable shape during a few days it is possible to track its mean rotation, which is in the sense of the atmosphere and the surface, and retrograde with respect to Earth. For different periods (orbits 251 – 254 and orbits 313 – 316) we measured a rotation period of -2.2 ± 0.2 days. This is close to the rotation period over the first 45 days of VEX observations (-2.5 days) [8] and similar to the rotation period of the north polar vortex (~ -3 days) observed by Pioneer Venus in 1979 [7].

We have selected 25 orbits to measure features' motions at the upper or lower cloud layers (see supplementary information). In 20 of these orbits we have obtained simultaneous measurements in both altitude levels. Representative examples are shown in Fig. 1. In most of these orbits the motions appear uncorrelated with the vortex morphology. There is an uncertainty with respect to the altitude level sampled by the 3.8-5.1 µm observations. A recent analysis of these thermal emissions places their altitude at 60 – 63 km [12]. Instead, a detailed study of VMC and VIRTIS images of the polar region locates the cloud top altitude of the UV markings at 63 – 69 km [11]. However, the morphology of the vortex as observed at 3.8 and 5.1 µm images correlates well with UV cloud features observed with the VIRTIS-M visible channel [8]. Additionally, the meridional profile of the zonal wind retrieved from tracking day-side UV features [15, 16] has the same behavior as the wind profile obtained in day and night-side observations at 3.8 and 5.1 µm (see supplementary information). In the upper and lower cloud the mean zonal wind at 75°S is about -40 ± 10 m s$^{-1}$ decreasing almost linearly to 0 ± 10 m s$^{-1}$ at the pole, implying an atmospheric rotation period of -2.9 ± 0.7 days compatible with the mean rotation period of the vortex. Thus the vortex seems to be an organized feature covering on average ~ 2,200 x 1,400 km² that is tied to the ambient zonal flow but whose structure is destroyed and reformed continuously.

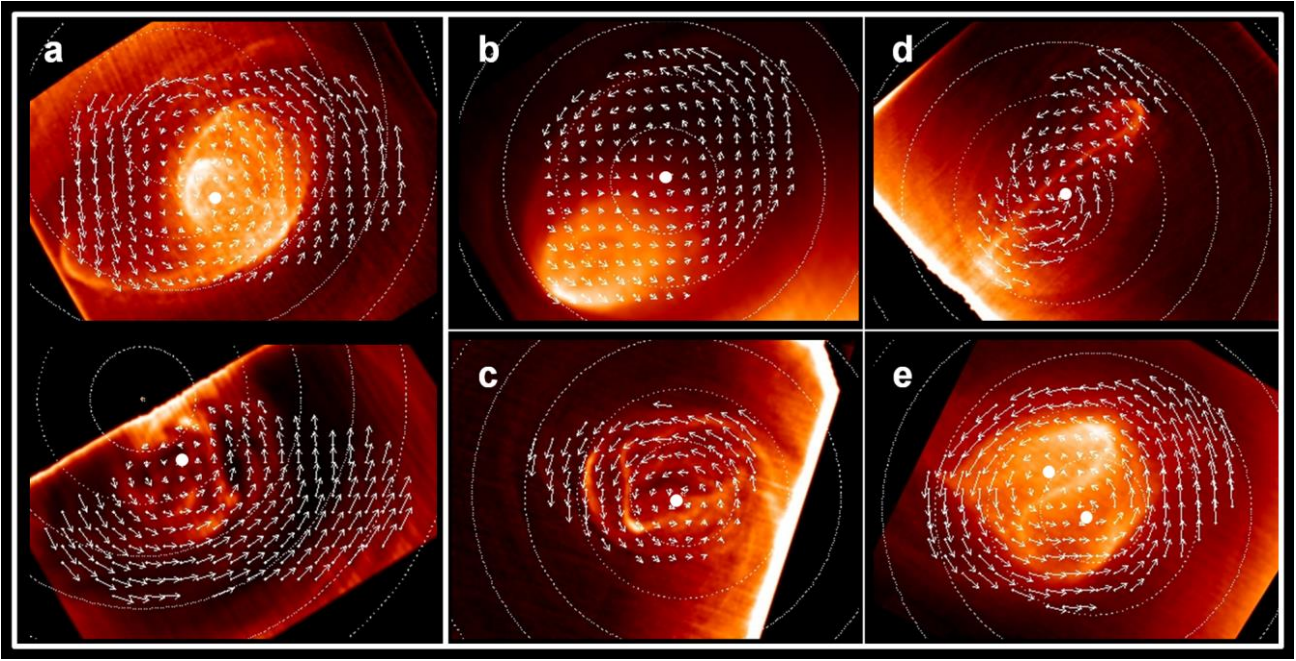

**FIGURE 1: Polar projections of vortex morphology and wind retrievals**. **a**, Upper (top) and lower (bottom) cloud levels on orbit 310. **b-e**, Upper cloud level on orbits 251, 355, 394 and 475, respectively. Latitude circles (dotted lines) appear every 5 degrees. The white points show the rotation center derived from the wind field. The largest vectors in each panel are 50 m s$^{-1}$ (a, upper cloud) and 46 m s$^{-1}$ (a, lower cloud), 37 m s$^{-1}$ (b), 40 m s$^{-1}$ (c), 38 m s$^{-1}$ (d), and 61 m s$^{-1}$ (e). Typical measurement errors are 4 m s$^{-1}$.

From the wind data we have calculated the relative vorticity and the streamlines associated to the instantaneous wind field (Fig. 2). Most of the vorticity maps show no outstanding features linked to the visible morphology of the polar vortex in neither of the two cloud layers. There are also no great differences between both height levels as for the values of vorticity or its spatial distribution. Vorticity maxima are found correlated with the vortex morphology only in some scarce cases (see panels A and C in Fig. 2). In the upper cloud where the vortex is observed with more contrast, bright and warm features generally locate surrounding the regions of maximum vorticity (see panels B and E in Figures 1 and 2, and Fig. 3). Interestingly, the streamlines (solid lines in Fig. 2) are not coupled with vorticity due to magnitude of the curvature term in the vorticity, which is of the same order as the wind shear's term in the polar region. In the upper level the streamlines show a closed circulation around a rotation centre in half of the analyzed cases. In the other half we find divergent (mass outflow) or convergent (mass inflow) circulations implying vertical velocities of up to 0.2 m s$^{-1}$. Panels D in figures 1 and 2 show a clear convergence of the wind field in the region from the pole to 85°S latitude that has an associated vertical velocity of $w \sim -0.16 \pm 0.01\, m\, s^{-1}$ and a volume flux of $\emptyset \sim -1.13 \pm 0.07 \times 10^{11}\, m^3\, s^{-1}$, assuming a value of 4km for the vertical scale height. In an alike divergent case we have estimated for a similar area $w \sim 0.05 \pm 0.01\, m\, s^{-1}$ and $\emptyset \sim 3.7 \pm 0.7 \times 10^{10}\, m^3\, s^{-1}$. In another case (panel E in Figures 1 and 2), there are two weak divergent regions near to two vorticity local maxima. In this case, the bright filament seen in the thermal image goes through the vorticity maxima "avoiding" them (Fig. 3).

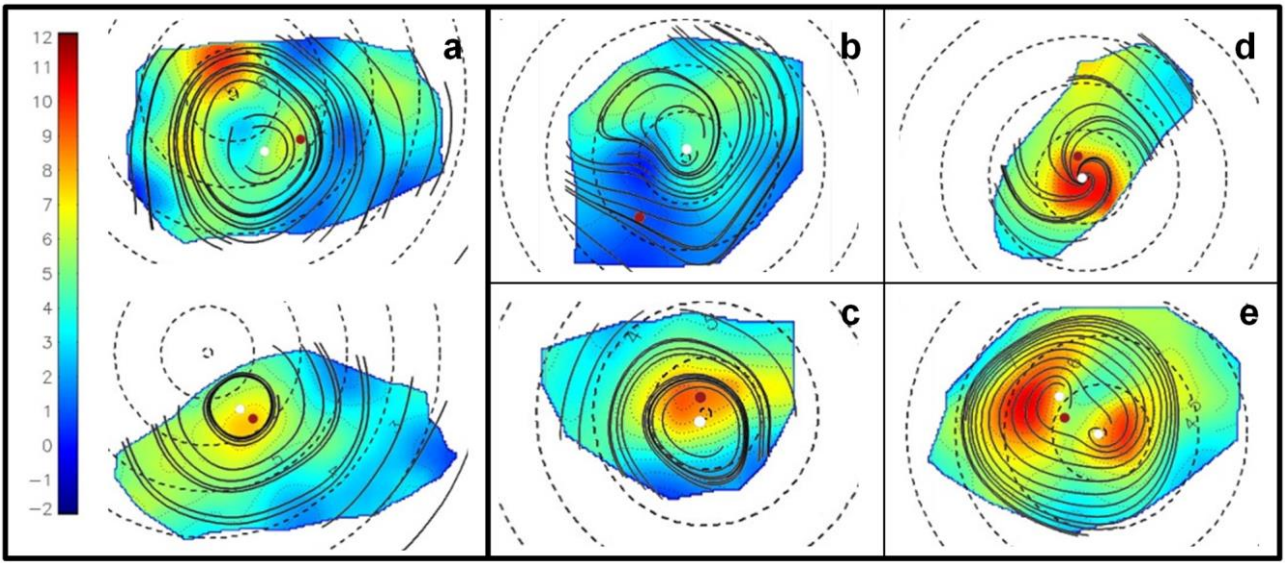

**FIGURE 2: Local relative vorticity maps (color coded, units are s$^{-1}$) and streamlines (continuous lines). a**, Upper (top) and lower (bottom) cloud levels on orbit 310. **b-e**, Upper cloud level on orbits 251, 355, 394 and 475, respectively. Latitude circles (dashed lines) appear every 5 degrees. The white points show the rotation center derived from the wind field and maroon points show the morphological center. Most of the vorticity maps look like panel (b), with no remarkable maxima. Vorticity errors are of the order of 2.5x10$^{-5}$ s$^{-1}$.

Figure 4 shows the rotation centers obtained from the streamlines patterns both at the upper and lower levels. In the lower layer we only observe the night-side of the planet and in those cases where the vortex extends into the day-side an extrapolation of the streamlines was necessary. We have tracked the position of the rotation center in different successive orbits and we have found that the vortex center is located at different positions in each altitude wandering erratically in both levels. Our analysis shows that the vortex center moves with speeds ranging from 3 to 16 m s$^{-1}$ in the lower layer (with a mean value of 8 ± 3 m s$^{-1}$) and from 0.4 to 16 m s$^{-1}$ in the upper layer (with a mean value of 5 ± 2 m s$^{-1}$). Part of this erratic behavior at the upper layer can be interpreted as a vortex precession around the pole [14]. However the motion seems to be more irregular in the lower level where the vortex center is displaced from the pole 4.0 ± 0.4° on average while in the upper layer the vortex stays longer near the pole and has an average displacement of 2.3 ± 0.4°. This behavior could support the Gierasch-Rossow-Williams mechanism for the origin of atmospheric superrotation, that involves vertical and meridional momentum transport by an axisymmetric Hadley circulation and horizontal transport by non-axisymmetric eddies generated by barotropic or baroclinic instabilities [17, 18, 19]. However, our data is not conclusive about solar tides, as it shows no or little dependence with local time (see supplementary material).

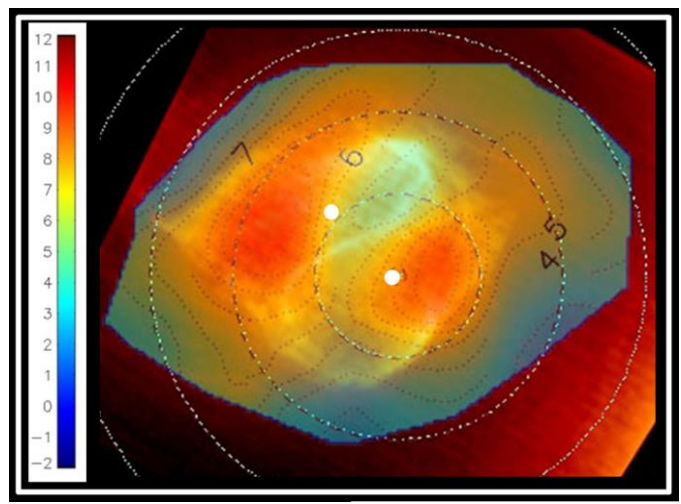

**FIGURE 3: Superimposed vorticity map (color coded, units are s$^{-1}$) and morphology.** This example corresponds to the upper cloud layer on orbit 475 (the morphology pattern can be seen in Fig. 1 panel E). Latitude circles appear every 5 degrees. Dotted contours represent vorticity isolines. The two white points show two centers of rotation as derived from the wind field. The high thermal emission features generally appear surrounding the patches of cyclonic vorticity maxima and can be centered in regions of nearly null vorticity. Vorticity errors are of the order of 2.5x10$^{-5}$ s$^{-1}$.

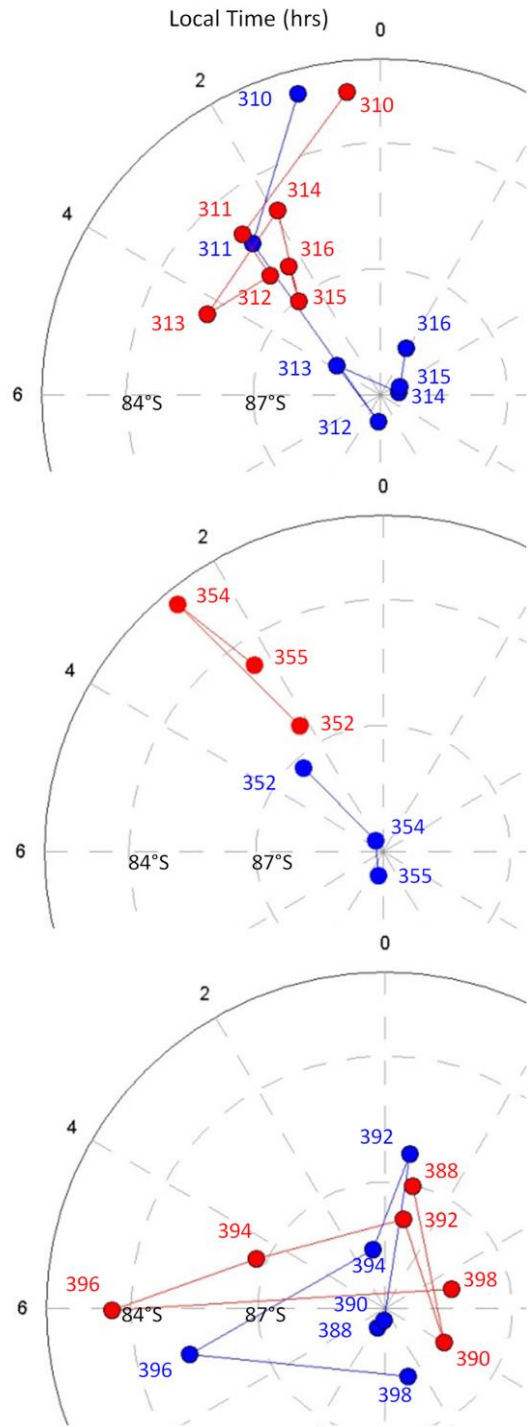

**FIGURE 4: Vortex's erratic wandering**. Vortex center of rotation for the lower (red) and the upper cloud layers (blue) in terms of latitude and local time (in hours) in a polar projection. Numbers correspond to particular orbits. A measurement of the vortex center was obtained in each single orbit, but we only plot here different short sequences of orbits and the trajectories run by the vortex center. These positions are accurate within 1° for the upper cloud and 4° for the lower cloud.

There are numerical models [21, 22] and a few laboratory experiments in rotating tanks [23] trying to explain the behavior of the Venus' polar vortex and its frequent dipolar shape. Those works are based on barotropic conditions that are not satisfied near the pole and fail to reproduce the variability of the vortex. In fact, from the mean meridional wind profiles derived from cloud tracking on VIRTIS images in the UV (380 nm, top upper cloud) and 980 nm (base upper cloud) [15], the vertical shear of the mean zonal wind in polar latitudes between altitudes ~ 65 and 60 km is <du/dz> ~ 5 ± 3 m s$^{-1}$ km$^{-1}$ . Our present measurements indicate that when averaged over several orbits there is low vertical shear of the zonal wind <du/dz> ~ 0.08 ± 0.06 m s$^{-1}$ km$^{-1}$, considering a large vertical extent from ~ 63 km (motions measured at 3.8 and 5.1 µm) to the lower cloud (~ 42 km at 1.74 µm). Although the mean vertical shear is small, our simultaneous single orbit measurements of cloud motions at both altitudes show that the vortex exhibits strong vertical shears that can be as high as |du/dz| ~ 0.8 ± 0.2 m s$^{-1}$ km$^{-1}$. On the other hand, temperature retrievals using the VeRa instrument onboard VEX [24, 25] show that a well developed tropopause exists only in the polar region with a peak inversion of ~ 16 K [24] at the cold collar at an altitude ~ 60 km (see supplementary information), with a meridional temperature gradient at this altitude of dT/dy ~ 6 ± 1 x 10$^{-3}$ K km$^{-1}$ [24]. The inversion separates two dynamical regimes. Above the tropopause in the upper cloud, the static stability is high $S_T = (dT/dz) + (g/C_P) \sim 10\ K\ km^{-1}$, but drops to zero downward to the lower cloud (~ 42 km). Consequently the Richardson number $Ri = \frac{gS_T}{T(du/dz)^2}$ changes abruptly in the vertical from ~ 5 to 20 (upward 60 km) to ~ 0 or negative values (downward 60 km) [17, 24, 26]. Although this trait is similar to other latitudes in Venus, the polar region is different in several aspects. First, the zonal wind profiles at the three altitude levels sensed by VIRTIS have a strong meridional shear <du/dy> ~ -0.017 ± 0.004 m s$^{-1}$ km$^{-1}$ . Likewise the meridional wind in the upper cloud measured in UV images shows small decreasing speeds poleward of 60° (equatorward they increase). When all these ingredients are put together, including the curvature term in the polar vorticity, the picture that emerges is that the polar vortex, resulting from barotropic or baroclinic instability depending on the rate of kinetic or potential energy supply, broadly extends between ~ 42 – 63 km and shows a high degree of variability. The source of this variability is possibly related to the location of the vortex in a baroclinic environment within a quasi-convective region embedded between the upper and lower cloud layers. Future dynamical models of the polar vortex should include this complex dynamical and thermal structure if they want to address its variability and unexpected wanderings across the pole.

**Methods Summary**

Images were processed to reduce noise and artifacts. All the images were polar projected with a spatial resolution of 10.5 km per pixel at the pole and contrast enhanced with a high-pass filter. We examined the motions of the lower atmosphere cloud features and the thermal emission from the top of the cloud deck assuming that these structures drift with the local wind. We used an image-correlation algorithm [27] that divides an image in small boxes and identifies the most probable match for each box in a second image. Boxes had typical sizes of 40x40 pixels. Since the images are typically noisy we supervised all the measurements looking at a map of the correlation function and the proposed identification. Half of the automatic measurements were rejected resulting in small standard deviations of wind vectors at small scales. Typical measurement errors were on the order of the image spatial resolution divided by the time difference between images and amounted to 4 m s$^{-1}$. Local vorticity, convergence and divergence were calculated transforming the zonal and meridional wind components into Cartesian coordinates centered at the pole. This automatically takes into account the curvature term associated to the polar circulation. For these calculations we interpolated the wind field in each image pair to a fix spatial resolution of 21 km and performed the spatial derivatives with a spatial scale of 525 km to avoid instabilities in the derivatives from using a small spatial step. Finally, streamlines were calculated by solving the advection of test particles in a constant wind field. Further details about error estimations are available in the supplementary material. The calibrated VIRTIS data used for this research is freely available at the Planetary Science Archive on http://www.rssd.esa.int/

**Acknowledgements** The authors declare no competing financial interests. We wish to thank Y. J. Lee and S. Tellmann for providing VeRa data. We gratefully acknowledge the work of the entire Venus Express team that allowed these data to be obtained. We wish to thank ESA to support the Venus Express mission, ASI (by the contract I/050/10/0), CNES and the other national space agencies supporting the VIRTIS instrument on board. J. Peralta acknowledges support from the Portuguese Foundation for Science and Technology (FCT, grant reference: SFRH/BPD/63036/2009). This work was supported by the Spanish MICIIN project AYA2009-10701 and AYA2012-36666 with FEDER support, Grupos Gobierno Vasco IT-464-07 and UPV/EHU UFI11/55.

**Author contributions**

I. G.-L. performed the image selection and wind measurements. R. H. designed the measurement software. A. S.-L. coordinated this research and with J. P. made theoretical interpretations. P.D. and G.P. have coordinated the observations as Principal Investigators of VIRTIS. All the authors discussed the results and commented on the manuscript.